\documentclass[12pt]{iopart}
\pdfoutput=1

\usepackage{graphicx}
\usepackage{subfig}
\usepackage{overpic}
\usepackage{epstopdf}
\usepackage{tikz}
\usetikzlibrary{arrows,shapes}

\expandafter\let\csname equation*\endcsname=\relax
\expandafter\let\csname endequation*\endcsname=\relax
\usepackage{amsmath}	
\usepackage{amssymb}    

\newcommand{\tikzcircle}[2][blue,fill=blue!50]{\tikz[baseline=-0.5ex]\draw[#1,radius=#2] (0,0) circle ;}%
\newcommand{\tikzcirclered}[2][red,fill=red!50]{\tikz[baseline=-0.5ex]\draw[#1,radius=#2] (0,0) circle ;}%
\begin{document}


\title[Enhanced Schwinger's pair production]{Enhanced Schwinger's pair production in many-center systems }

\author{Fran\c{c}ois Fillion-Gourdeau$^{1}$, Emmanuel Lorin$^{2,1}$, Andr\'{e} D. Bandrauk$^{3,1}$}

\address{$^1$ Centre de Recherches Math\'{e}matiques, Universit\'{e} de Montr\'{e}al, Montr\'{e}al, Canada, H3T 1J4}
\address{$^2$ School of Mathematics and Statistics, Carleton University, Ottawa, Canada, K1S 5B6}
\address{$^3$ Laboratoire de chimie th\'{e}orique, Facult\'{e} des Sciences, Universit\'{e} de Sherbrooke, Sherbrooke, Canada, J1K 2R1}

\ead{filliong@CRM.UMontreal.ca}

\begin{abstract}
Electron-positron pair production is considered for many-center systems with multiple bare nuclei immersed in a constant electric field. It is shown that there are two distinct regimes where the pair production rate is enhanced. At small interatomic distance, the effective charge of the nuclei approaches the critical charge where the ground state dives into the negative continuum. This facilitates the transition from the negative to the positive energy states, which in turns, increases the pair production rate. At larger atomic distance, the enhancement is due to the crossing of resonances and the pair production proceeds by the Resonantly Enhanced Pair Production (REPP) mechanism. These processes are studied within a simple one-dimensional model. A numerical method is developed to evaluate the transmission coefficient in relativistic quantum mechanics, which is required in the calculation of the pair production rate. The latter is evaluated for systems with many (up to five) nuclei. It is shown that the production rate for many-center systems can reach a few orders of magnitude above Schwinger's tunnelling result in a static field.
\end{abstract}

\submitto{\jpb}

\section{Introduction}

In the last few years, there has been a surge of interests for Schwinger's mechanism and its variants, motivated mostly by new technological advances in laser physics, which will allow to reach electromagnetic field intensities exceeding $I \sim 10^{23}$ W/cm$^{2}$ in the near future \cite{RevModPhys.78.309}. At these intensity levels, relativistic effects start to be important \cite{Salamin200641} and new Quantum Electrodynamical effects, such as vacuum polarization and the production of electron-positron pairs, may be observed \cite{RevModPhys.84.1177}. In particular, there is a possibility of studying Schwinger's process, which has eluded experimental confirmation for the last fifty years.  

Schwinger's mechanism concerns pair production from a constant electric field \cite{PhysRev.82.664}: in the Dirac interpretation, this is understood as a tunnelling process for an electron in the negative energy sea to a positive energy state. The probability to produce a pair is then given by \cite{PhysRev.82.664,Itzykson:1980rh}
\begin{eqnarray}
P_{S} \sim e^{- \frac{\pi m^{2} c^{4}}{c|e|\hbar E}},
\end{eqnarray}
where $m$ is the electron mass, $c$ the speed of light, $|e|$ the absolute value of the electron charge and $E$ the electric field strength. Thus, it can be concluded that there is an important amount of pairs produced when $P_{S} \sim 1$, which occurs when the electric field reaches Schwinger's electric field strength given by
\begin{eqnarray}
E_{S} = \frac{m^{2} c^{3}}{e\hbar} \approx 1.3 \times 10^{18} \; \mbox{V/m}.
\end{eqnarray}
Although the field strengths attained in the new ultra-intense laser regime are still far from Schwinger's ``critical'' field strength $E_{S}$ (corresponding to an intensity of $I \approx 2.24 \times 10^{29}$ W/cm$^{2}$), it may still be possible to observe this Quantum Electrodynamic effect by using other external field configurations.

To reach this goal, many scenarios has been studied such as counterpropagating lasers \cite{PhysRevD.2.1191,PROP:PROP19770250111,Nikishov1973,PhysRevE.66.016502}, counterpropagating lasers with space dependence \cite{PhysRevLett.102.080402}, the interaction of laser field with heavy nuclei \cite{PhysRevA.67.063407,PhysRevLett.100.010403,PhysRevLett.91.223601,PhysRevLett.103.170403,
0953-4075-31-8-012,PhysRevA.76.012107} and the combination of rapidly and slowly varying fields \cite{PhysRevA.85.033408,PhysRevD.80.111301}. The effect of the temporal laser pulse shape has also been investigated \cite{PhysRevD.70.053013,PhysRevLett.104.250402,PhysRevD.83.065028}. Most of these ideas take advantage of some of the following effects:
\begin{itemize}
	\item Time-dependence of the laser field, which effectively shift energy levels by $\hbar \omega$ (where $\omega$ is the laser frequency).
	\item The presence of a nucleus, which has a ground state at lower energy level than in vacuum.
	\item Volume effects: in a laser field, pairs are produced in a bulk. Thus the small production rate (per unit volume) can be compensated by a large enough volume.
	\item Cascading effects: once a pair is created (seed), the electron and positron can emit hard or virtual photons which can decay into other pairs.  
\end{itemize}
Another important mechanism is the so-called trident process where an incident electron is propagated through the strong field and emits a virtual photon \cite{PhysRevLett.106.020404}. The latter then decays into an electron-positron pair. This was used in the celebrated SLAC experiment to produce an observable number of pairs in the collision of an electron with a laser pulse \cite{PhysRevLett.79.1626}. 

In this article, electron-pair production from a many-center system immersed in a constant electric field is considered. More precisely, the system is composed of fully ionized atomic nuclei while the electric field could be supplied by a slowly varying (low frequency) counterpropagating laser field \cite{springerlink:10.1007/978-3-642-28726-8_2,PhysRevLett.89.193001} (non-adiabatic effect are ignored, although they may be included as in \cite{PhysRevA.86.032118}). Both the nuclei and laser field are treated in the external field approximation and are described by a classical field, which is allowed when the number of photons is large such that quantum fluctuations can be neglected.

The main result of this article is that when many nuclei are considered, there are two distinct regimes where the production of pairs is enhanced, in comparison to Schwinger's process:
\begin{itemize}
\item Small interatomic distance: the production of pair proceeds by the Effective Charge Enhancement Pair Production (ECEPP) process.
\item Larger interatomic distance: the production of pair proceeds by the Resonantly Enhanced Pair Production (REPP) process \cite{PhysRevLett.110.013002,1742-6596-414-1-012013}.
\end{itemize}
Both of these mechanisms are described in more details in Section \ref{sec:ppp}. They are studied theoretically in a one-dimensional model where the nuclei are represented by delta function potential wells. This model is very simple but it contains the main physical features required to understand the variations in the pair production rate. Other 1-D models have been considered in the literature \cite{PhysRevD.83.065028,PhysRevLett.108.030401,PhysRevLett.104.250402,PhysRevA.80.062105} while an analytical approach to the relativistic ionization of a 3D two-center system can be found in \cite{0953-4075-42-17-171003}.

This article is organized as follows. In Section \ref{sec:ppp}, we present the two mechanisms (REPP and ECEPP) in details. In Section \ref{sec:pair_prod}, the procedure to evaluate the pair production rate in inhomogeneous strong external field is described. The model used to characterize the nuclei is given in Section \ref{sec:model}. The numerical method used to solve the Dirac equation and the transmission-reflection problem is presented in Section \ref{sec:num_meth}. Numerical results concerning the position of resonances and the pair production rate in many-center systems can be found in Section \ref{sec:res}. We conclude in Section \ref{sec:conclu}. Also, units in which $\hbar = c  = m = 1$ (where $m$ is the electron mass) and $e=\sqrt{\alpha}$ are utilized in most numerical calculations. In this case, the unit length is $l_{\rm u} = \hbar/(mc) \sim 3.86159 \times 10^{-13}$~m (0.386 pm) while the unit time is $t_{\rm u} = \hbar / (mc^{2}) \sim 1.2880885 \times 10^{-21}$~s (1.288 zs), as compared to atomic units: $l_{\rm a.u.} = 0.052$ nm and $t_{\rm a.u.} = 24 \times 10^{-18}$ s (24 as).

\section{Two mechanisms}
\label{sec:ppp}

In this section, the two mechanisms by which pair production is enhanced in many-center systems are described. It should also be noted that similar ideas, for the one-center system ECEPP, were given in \cite{0953-4075-31-8-012}.

\begin{figure}
\subfloat[]{\includegraphics[width=0.5\textwidth]{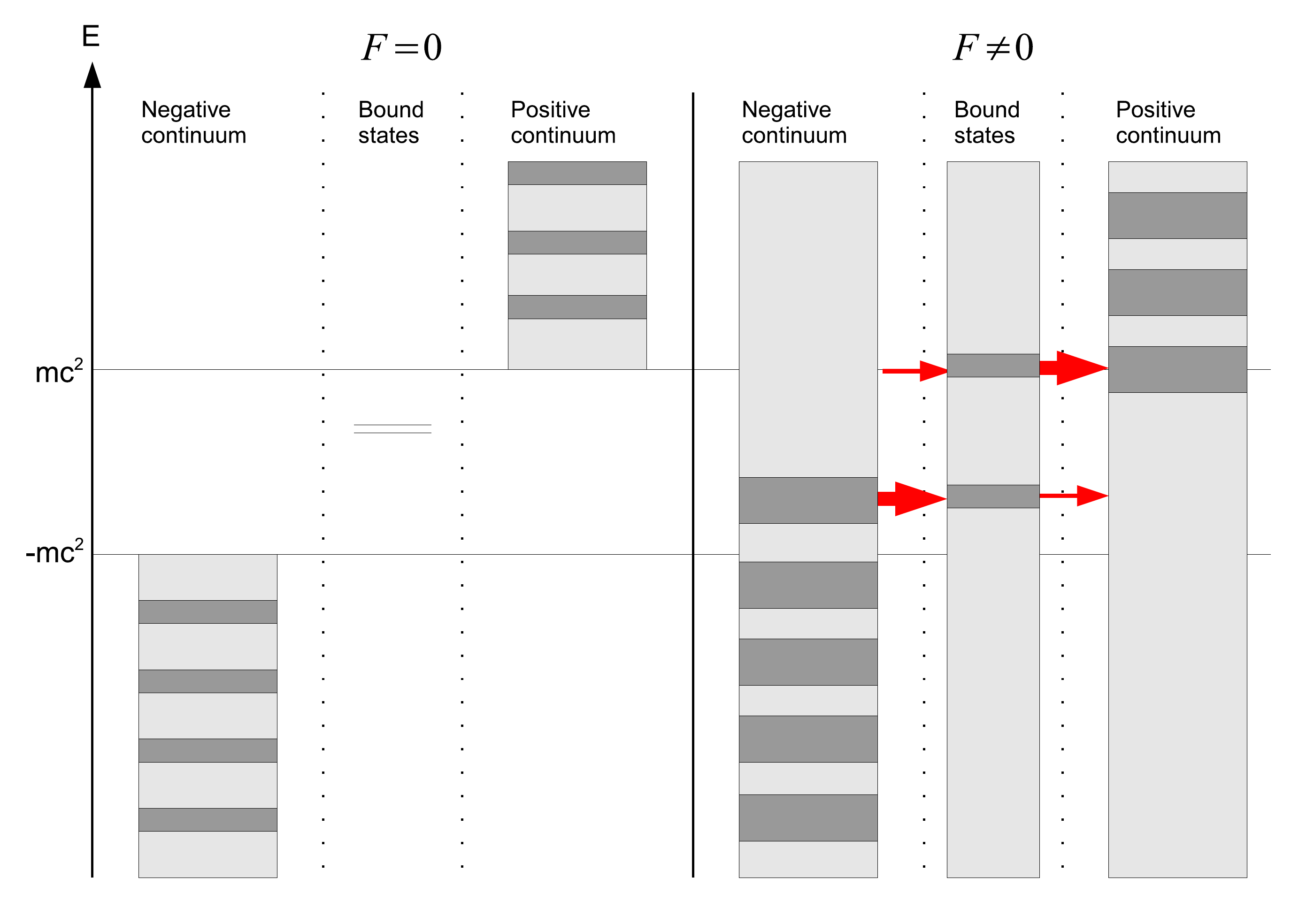}}
\subfloat[]{\includegraphics[width=0.5\textwidth]{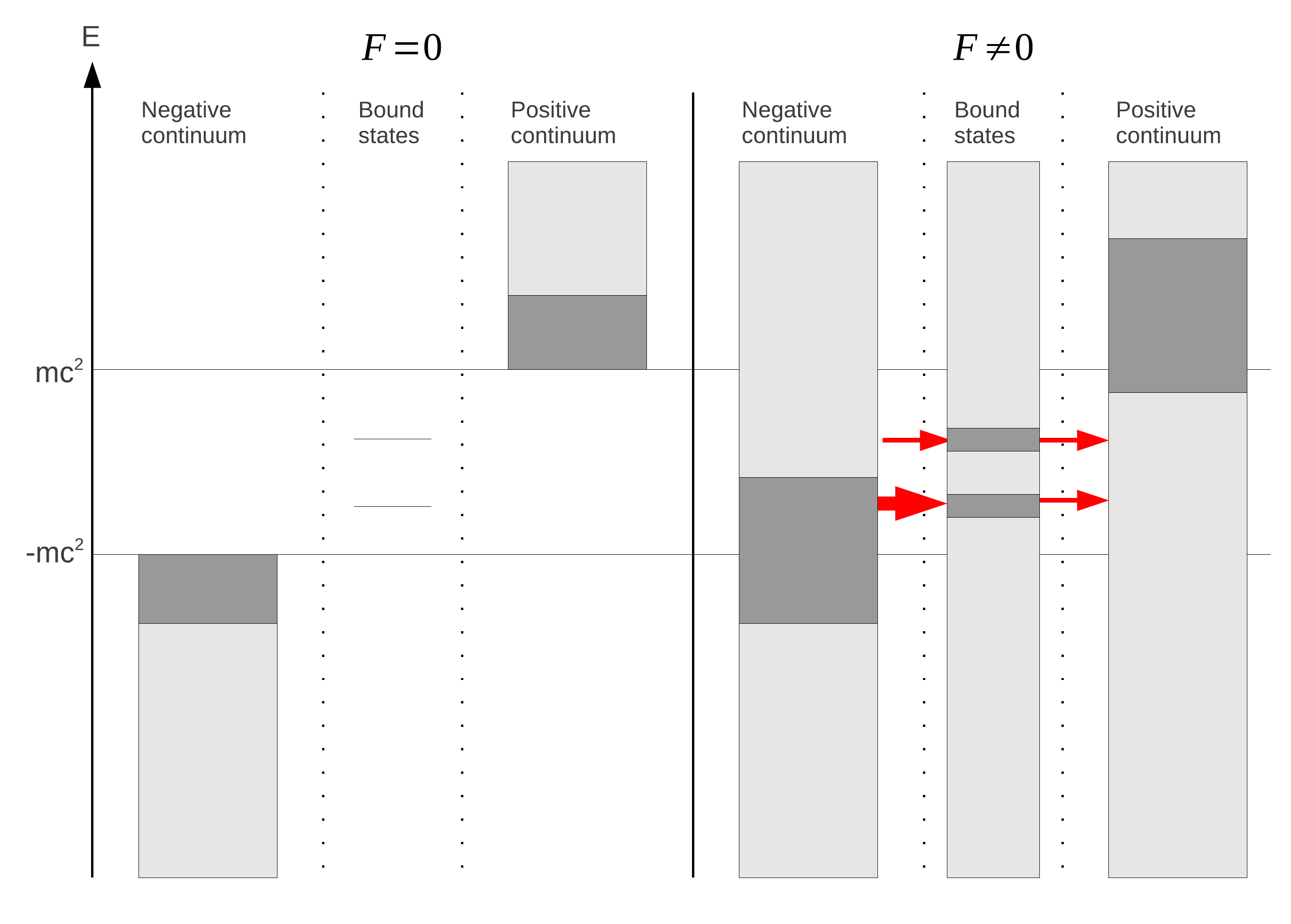}}
\caption{Description of the (a) REPP and (b) ECEPP mechanisms for the two-center system. The dark gray regions are the position of resonances, which have a high density of states, while the light gray regions correspond to accessible energies which have a lower density of state. In (a), the interatomic distance is large and the ground and first excited states are quasi-degenerate. These two states are Stark shifted in the presence of the field. In (b), the interatomic distance is small such that the ground state has a low energy, close to $-mc^{2}$. In that limit, the Stark shift is less important than for large $R$.}
\label{fig:REPP}
\end{figure}

\subsection{REPP}

The REPP mechanism \cite{PhysRevLett.110.013002} is very similar to Charge Resonance Enhanced Ionization (CREI), which describes the non-relativistic non-perturbative ionization of diatomic molecules \cite{PhysRevA.52.R2511,springerlink:10.1007/978-3-642-28726-8_2}. It proceeds when resonances from the negative energy states crosses with resonances from the positive energy states: at the crossing, a transition between them is possible and this enhances pair production. It was shown in  \cite{1751-8121-45-21-215304} for the two-center case that the resonances in the negative energy states, which are related to the backscattering on the potential wells, are Stark shifted to higher energies as the electric field and the interatomic distance are increased. On the other hand, it is well-known that the ground state and the first excited state resonances, at large interatomic distance, are Stark-shifted by $\Delta E \sim \mp RF/2$ (for the case of two nuclei). They also gain an imaginary part because in the electric field, they become unstable resonances coupled to the continuum. Thus, for large enough $R$ and $F$, the ground state resonance can cross with negative energy resonances. Therefore, REPP is important at the interatomic distance and field strength where this crossing occurs, resulting in a peak in the pair production rate (this will be shown explicitly in the next sections). The REPP is illustrated in Fig. \ref{fig:REPP}.

\subsection{ECEPP}

The second mechanism occurs at small interatomic distance, when the nuclei are close to each other. In this case, without the constant electric field, the ground state approaches $-mc^{2}$ due to the increasing effective charge which becomes $Z_{\rm eff} \approx NZ$ (for $Z$ the electric charge of one nucleus and $N$ the number of nuclei). It is well-known that when $Z_{\rm eff} \geq Z_{\rm crit}$, where $Z_{\rm crit}$ is the critical charge where the ground state energy reaches $E_{1s} = -mc^{2}$, pair production starts to be possible.  This pair creation mechanism was studied extensively in relativistic heavy ion collisions \cite{PhysRevA.24.103,Baur20071} and was detected experimentally \cite{PhysRevLett.56.444,PhysRevLett.54.1761,Clemente198441}. When the electric field is turned on, the energy shift of the ground state resonance (the ground state becomes an unstable state in the electric field) is unimportant because in that regime, the Coulomb potential dominates over the Stark shift. However, because the effective charge approaches the critical value and the ground state is closer to $-mc^{2}$, it is easier for an electron of the negative energy states to tunnel to positive energy states via the ground state. Therefore, the presence of the electric field allows pair production even if the effective charge does not reach the critical value where the ground state dives into the negative energy continuum. Moreover, the presence of the ground state resonance facilitates the tunnelling effect and this enhances pair production. Finally, the electric field reduces the Pauli blocking in the ground state because the latter can be ionized. After ionization, another electron can tunnel from the negative energy states, creating a flux of electron and positron moving in opposite direction. The ECCEPP mechanism is depicted in Fig. \ref{fig:REPP}.

\section{Pair production rate}
\label{sec:pair_prod}

The rate of producing electron-positron pairs from a time-dependent strong external field is an observable that has been studied extensively for many applications since the pioneering work of Nishikov \cite{Nikishov1964}. Starting from the definition of the average number of pairs produced, which is given by $\langle n \rangle = \sum_{n} n P_{n}$, where $P_{n}$ is the probability of producing $n$ pairs, and assuming the external field vanishes asymptotically at times $t=\pm \infty$, there exists a direct link between the retarded solution of the Dirac equation (including the external field) and $\langle n \rangle$ \cite{Gelis2006135,PhysRevC.71.024904}. To obtain $\langle n \rangle$ explicitly, it is required that all negative energy states are propagated in time and are projected over all positive energy states. 

In the case of a time-independent external field, a similar reasoning can be used. First, the governing equation is the following Dirac equation (in 1-D):
\begin{equation}
 E\psi(x) = \left[ -ic \alpha \partial_{x} + \beta mc^{2} + V(x) + A_{0}(x)  \right] \psi(x), \;\; x \in \mathbb{R},
\end{equation}
where $\psi$ is a bi-spinor, $V$ is the potential of the nuclei, $A_{0}(x)$ is the potential of the laser field, $m$ is the electron mass and $E$ is the energy. The Dirac matrices are given by the following representation:
\begin{eqnarray}
 \alpha = \sigma_{z} \;\; \mbox{and} \;\; \beta = \sigma_{x},
\end{eqnarray}
where $\sigma_{x,y,z}$ are the usual Pauli matrices. This representation is convenient as it yields simple equations. It can be shown easily that they obey the appropriate anticommutation relations. 

To relate pair production to a solution of this equation, one assumes the vanishing of the field at $x=\pm \infty$ (thus, we have $V(x) \xrightarrow{x \rightarrow \pm \infty} 0$ and $A_{0}(x) \xrightarrow{x \rightarrow \pm \infty} C$, with $C$ a constant). Then, it is possible to define ``asymptotic states'' at $x=\mp \infty$: in these regions, the particles are free and there is an unambiguous separation between the negative and positive energy solutions. Using this field configuration, it can be shown that the production rate (per unit volume and time $t$) is given by  \cite{1977mgm..conf..459D,PhysRevD.38.348,1402-4896-23-6-002,PhysRevD.82.025015,PhysRevD.80.065010} 
\begin{eqnarray}
\label{eq:pair_prod_master}
\frac{d \langle n \rangle}{dtdE} &=& \frac{1}{2\pi} |A(E)|^{2}, \;\; E\in \Omega_{\rm Klein}, \\
\frac{d \langle n \rangle}{dt} &=& \frac{1}{2\pi} \int_{\Omega_{\rm Klein}}dE |A(E)|^{2},
\end{eqnarray}
where $A$ is the coefficient of the positive energy solution propagating towards $x=+\infty$, at the right of the potential (see Fig. \ref{fig:pair_prod_calc}). The energy $E$ is in the Klein region $\Omega_{\rm Klein}$ where a transition from a negative to a positive energy state is possible (the blue region in Fig. \ref{fig:pair_prod_calc}). Thus, the calculation of pair production reduces to a transmission-reflection problem where the incident, reflected and transmitted waves are given respectively by:
\begin{eqnarray}
\begin{cases}
\psi_{\rm inc.}(x) = v(p)e^{ip(E)x} & x\in (\infty,-L] ,\\
\psi_{\rm ref.}(x) = Bv(-p)e^{-ip(E)x}  & x\in (\infty,-L],\\
\psi_{\rm trans.}(x) = Au(k)e^{ik(E)x}  & x\in [L,\infty),
\end{cases}
\end{eqnarray}
where we assume that the external field is non-zero only for $x\in (-L,L)$. Here, $p,k$ are plane wave momenta while $u,v$ are the positive and negative energy free spinors: their explicit expression will be given below as their explicit expression depends on the external field considered.

To solve the transmission-reflection problem, the continuity of the wave function at $x=\pm L$ is imposed. Then, we obtain the following system of equation:
\begin{eqnarray}
v(p)e^{ipL} + B v(-p)e^{-ipL} =  \psi(-L), \\
\psi(L) = Au(k)e^{ikL} .
\end{eqnarray}
where $\psi(x)$ is the solution of the Dirac equation for $ x\in [-L,L]$. These conditions allows to determine the transmission coefficient $A$ once the wave function $\psi$ is determined. However, for numerical computations, it is actually more convenient to write the last equations as
\begin{eqnarray}
\label{eq:trans1}
\frac{v(p)e^{ipL} + B v(-p)e^{-ipL}}{A} =  \tilde{\psi}(-L), \\
\label{eq:trans2}
\tilde{\psi}(L) = u(k)e^{ikL},
\end{eqnarray}
where we defined $\tilde{\psi} := \psi/A$. Thus, the RHS of Eq. \eqref{eq:trans2} can be used as an initial condition. Then, the transmission coefficient will be given by 
\begin{eqnarray}
\label{eq:A_coeff_def}
A(E) = \frac{v_{2}(-p)v_{1}(p) - v_{1}(-p)v_{2}(p)}{v_{2}(-p)\tilde{\psi}_{1}(-L) - v_{1}(-p)\tilde{\psi}_{2}(-L)}e^{2ipL}
\end{eqnarray}
The numerical determination of $\tilde{\psi}$ is the subject of Section \ref{sec:num_meth}.

\section{Simple model}
\label{sec:model}

To have an accurate approximation of a nuclei, in the external field approximation, the nuclei should be modelled by 3-D Coulomb potentials (or other Coulomb-like potentials which include the charge distribution in the nucleus). Performing the calculation of the pair production rate in this case is a challenging task because it requires a numerical solution of the Dirac equation. For 3-D calculations, the latter requires a lot of computational resources and the numerical methods has to be carefully designed to circumvent problems related to fermion doubling \cite{PhysRevD.26.468} and variational collapse \cite{QUA:QUA560250112}. For these reasons, a simplified approach is used in this article where a 1-D model is utilized. However, it has been shown that in 1-D, this potential does not support bound states \cite{0305-4470-36-18-303}. The mechanisms considered in this article depends crucially on the position of bound states and their modifications in the field. Therefore, the nuclei will be modelled by delta function potential wells as 
\begin{eqnarray}
\label{eq:pot_2dirac_free}
 V(x) = -\sum_{i=1}^{N_{p}} g  \delta(x-R_{i})  , \;\; x \in \mathbb{R},
\end{eqnarray}
where $g$ is the strength of the Dirac delta potential well (physically, this parameter is very similar to the electric charge of the nucleus), $N_{p}$ is the number of nuclei and $R_{i}$ is their positions. It is well-known that these Dirac delta potential wells support bound state solutions \cite{1751-8121-45-21-215304}. Throughout this work, the potential wells strength is fixed to $g=0.8$ as this value reproduces the ground state energy of the 1$s$ orbital of the $U^{91+}$ atom. 

The position of the potential wells $R_{i}$ are chosen such that the internuclei distance $R$ is always the same. Thus, the following cases will be considered:
\begin{enumerate}
	\item No nucleus: $V(x)=0$
	\item Single nucleus: $V(x)=-g\delta(x)$
	\item Two nuclei: $V(x) = -g \delta(x-R/2) -g \delta (x+R/2)$
	\item Three nuclei: $V(x) = -g \delta(x-R)- g \delta(x) -g \delta (x+R) $
	\item Four nuclei: $V(x) = -g \delta(x-3R/2) -g \delta (x-R/2) -g \delta (x+R/2)-g \delta (x+3R/2)$
	\item Five nuclei: $V(x) =-g \delta(x-2R) -g \delta(x-R)- g \delta(x) -g \delta (x+R) -g \delta(x+2R)$
\end{enumerate}
The no-nucleus case corresponds to Schwinger's mechanism with a field having a finite extent (in numerical calculations, the region where the field is non-zero is chosen large enough such that boundary effects are negligible). The other ones physically represent linear clusters of heavy nuclei separated by a constant internuclei distance $R$.     

As in non-relativistic quantum mechanics \cite{albeverio2005solvable}, this type of point potential can be characterized by boundary conditions at the potential well positions: outside of these points, the potential is solely due to $A_{0}$. Finding these boundary conditions is a subtle problem because it involves product of distributions, which are not well-defined mathematically in the usual approach. This occurs because in the presence of a point interaction (delta function potential well), the wave function has a jump discontinuity and thus, has the form $\psi(x) \sim \sum_{i}\theta(x-R_{i})f(x)$ where $\theta(x)$ is the Heaviside function and $f(x)$ is the solution of the Dirac equation\footnote{In non-relativistic quantum mechanics, the discontinuity is in the derivative of the wave function. Thus, it is not problematic as in the relativistic case.}. Thus, the Dirac equation has terms behaving like $ \sum_{i}\delta(x-R_{i})\theta(x-R_{i})f(x)$, which are not well-defined mathematically (not uniquely defined). It is however possible to give a mathematical meaning to these product of distributions by using Colombeau's theory of generalized function \cite{Colombeau198396,Colombeau1990}. Using this theory, along with charge conjugation invariance of the Dirac equation and properties of self-adjoint extensions, it is possible to single out one boundary condition. It is given by \cite{1751-8121-45-21-215304} 
\begin{eqnarray}
 \psi( R_{i}^{+}) &=& G \psi( R_{i}^{-}),
\end{eqnarray}
where the transfer matrix is 
\begin{eqnarray}
G:= \left[ 1+\frac{g^{2}}{4c^{2}} \right]^{-1}
\begin{bmatrix}
 1- \frac{g^{2}}{4c^{2}} + i\frac{g}{c} & 0 \\
 0 &  1- \frac{g^{2}}{4c^{2}} - i\frac{g}{c} 
\end{bmatrix},
\end{eqnarray}
and where we defined $\psi( R_{i}^{+}) = \lim_{\epsilon \rightarrow 0} \psi( R_{i} + \epsilon)$ and $\psi( R_{i}^{-}) = \lim_{\epsilon \rightarrow 0} \psi(R_{i} - \epsilon)$.

\begin{figure}
\centering
\includegraphics[width=0.7\textwidth]{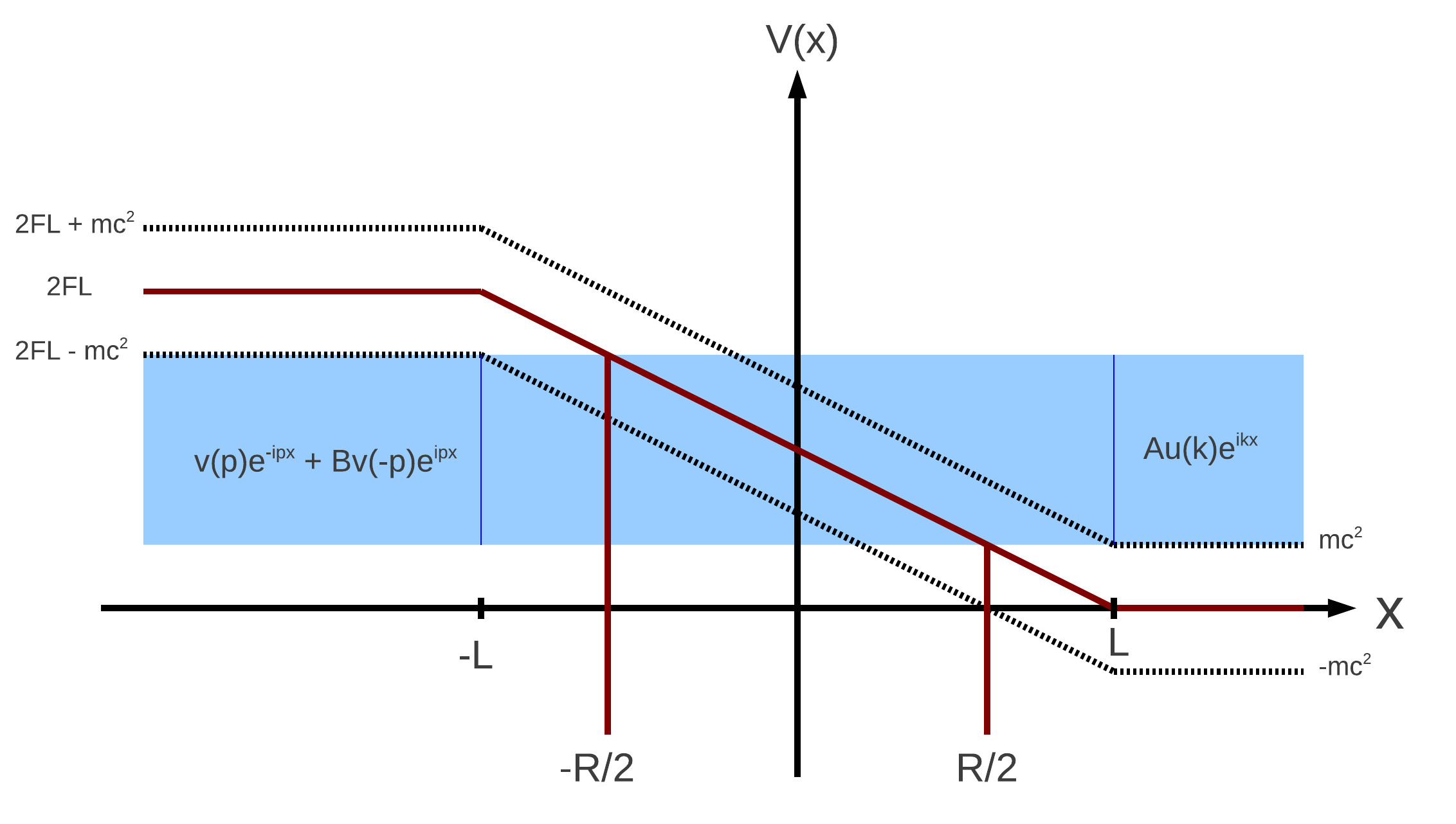}
\caption{Simple model to study pair production for many-center systems (here, the two-center system is shown). The electric field has a finite extent in space. The pair production rate calculation reduces to a transmission-reflection problem: the incoming, reflected and transmitted wave function are given in the figure. In blue is the Klein region where it is possible to have a transition from a negative to positive energy states. }
\label{fig:pair_prod_calc}
\end{figure}

The external field considered in this work corresponds to a constant electric field and is given by (see Fig. \ref{fig:pair_prod_calc})
\begin{eqnarray}
A_{0}(x) = 
\begin{cases}
2FL & \mbox{for} \; x \in (-\infty,-L] \\
-F(x-L) & \mbox{for} \; x \in (-L,L) \\
0 & \mbox{for} \; x \in [L,\infty) 
\end{cases}
\label{eq:elec_pot}
\end{eqnarray}
where $F$ is the field strength. It can be verified easily, by using a gauge where the vector potential is zero, that the electric field is constant over a length $2L$ and vanishes outside the interval $[-L,L]$. The Klein region for this external field is given by $\Omega_{\rm Klein} = [2FL-mc^{2},mc^{2}]$.

To compute the pair production rate, we also need the expression of the negative and positive energy free spinors in the regions where the electric field is zero. These quantities can be computed by seeking  plane wave solutions. This yields
\begin{eqnarray}
 u(k) &=& \frac{1}{\sqrt{2E}}
\begin{bmatrix}
 \sqrt{E+ck(E)} \\
 \sqrt{E-ck(E)}
\end{bmatrix}, \\
%
%
v(p) &=& \frac{1}{\sqrt{2(E-2FL)}}
\begin{bmatrix}
 \sqrt{(E-2FL)+cp(E)} \\
- \sqrt{(E-2FL)-cp(E)}
\end{bmatrix},
\end{eqnarray}
where $k(E)=\frac{1}{c}\sqrt{E^{2}-m^{2}c^{4}}$ and $p(E)=\frac{1}{c}\sqrt{(E-2FL)^{2}-m^{2}c^{4}}$.

\section{Numerical method for the computation of transmission coefficients}
\label{sec:num_meth}

The calculation of the transmission coefficient involves the solution of the time-independent Dirac equation. For a constant electric field, there exists a well-known analytical solution written in terms of parabolic cylinder functions \cite{sauter1931}. The pair production rate in \cite{PhysRevLett.110.013002} was evaluated by using this solution (it is shown in \ref{app:an_sol}). However, for smaller value of the electric field, the numerical evaluation of the parabolic cylinder function becomes problematic because the asymptotic expansions used to evaluate these special functions converge slowly in that limit. It is actually much more efficient to evaluate the wave function by a direct numerical solution of the Dirac equation. The numerical method used is now described.

The starting equation is the 1-D Dirac equation given by
\begin{eqnarray}
E \psi(x) = \left[ic \sigma_{z} \partial_{x}  + \sigma_{x} mc^{2} -F(x-L) \right]\psi(x), \;\; x \in [-L,L].
\end{eqnarray}
This gives the solution for the wave function when the electric field is non-zero (the inclusion of the potential wells is discussed below). Multiplying this equation on the left by $i\sigma_{z}/c$, we get
\begin{eqnarray}
\partial_{x} \psi(x) = \frac{1}{c}\left[ -\sigma_{y} mc^{2} - i\sigma_{z}\left[F(x-L) +E \right] \right]\psi(x).
\end{eqnarray}
The solution to this equation can be written formally as
\begin{eqnarray}
\psi(x_{ f}) &=& \mathcal{P} \exp \left\{ \frac{1}{c} \int_{x_{ i}}^{x_{ f}}dy\left[ -\sigma_{y} mc^{2} - i\sigma_{z}\left[F(y-L) +E \right] \right] \right\} \psi(x_{ i}) \\
&=& U(x_{i},x_{f} )\psi(x_{ i}),
\end{eqnarray}
where $x_{i,f}$ are the initial and final coordinates, $U$ is the space evolution operator and $\mathcal{P}$ represents the path-ordered exponential. The latter is required as the operator in the exponential does not commute with itself at different space position\footnote{We define $G(x):=-\sigma_{y} mc^{2} - i\sigma_{z}\left[F(y-L) +E\right]$. Then, it is easy to show that the commutator obeys $[G(x),G(y)] \neq 0$.}. Using the properties of this operator, we can write the last equation as
\begin{eqnarray}
\psi(x_{ f}) 
&=& U(x_{f},x_{n} )U(x_{i},x_{n-1} ) \cdots U(x_{2},x_{i})\psi(x_{ i}),
\end{eqnarray}
where we partitioned the space interval $[x_{i},x_{f}]$ into $n$ subintervals of size $\delta x = x_{j+1}-x_{j}$. Using the result of \cite{0305-4470-23-24-019}, it can be shown that the path ordered exponential can be approximated by
\begin{eqnarray}
U(x_{j},x_{j+1}) &=& \exp \left\{ \frac{1}{c} \int_{x_{ j}}^{x_{ j+1}}dy\left[ -\sigma_{y} mc^{2} - i\sigma_{z}\left[F(y-L) +E \right] \right] \right\} \nonumber \\
&&+ O(\delta x^{3}).
\end{eqnarray}
In other words, neglecting the path ordering results in an error which scales like $O(\delta x^{3})$ and the outcome is a second order numerical scheme. The integral in the exponential can be evaluated explicitly as 
\begin{eqnarray}
U(x_{j},x_{j+1}) &=& \exp \left\{  -\sigma_{y} mc \delta x + i\frac{\sigma_{z}\delta x}{c}\left[FL-E   - F\bar{x}_{j}  \right]\right\} \nonumber \\
&&+ O(\delta x^{3}),
\end{eqnarray}
where $\bar{x}_{j}:=\frac{x_{j}+x_{j+1}}{2}$ is the average position. Finally, written in this form, the exponential can be evaluated exactly using the properties of Pauli matrices. We find that 
\begin{eqnarray}
U(x_{j},x_{j+1}) &\approx& 
\begin{cases}
\mathbb{I}_{2} \cos(D) + i\sigma_{z} B \mathrm{sinc}(D) - \sigma_{y}C \mathrm{sinc}(D), & E \in \mathcal{D}_{1} \\
\mathbb{I}_{2} \cosh(\tilde{D}) + i\sigma_{z} B \mathrm{sinhc}(\tilde{D}) - \sigma_{y}C \mathrm{sinhc}(\tilde{D}), & E \in \mathcal{D}_{2}
\end{cases}
\end{eqnarray}
where $\mathrm{sinc}(z):=\sin(z)/z$ is the cardinal sine function and $\mathrm{sinhc}:=\sinh(z)/z$ is the cardinal hyperbolic sine function. Also, we defined
\begin{eqnarray}
B &:=& \cfrac{\delta x}{c} (FL-E-F\bar{x}_{j}) ,\\
C &:=& \cfrac{\delta x}{c} mc^{2} ,\\
D &:=& \sqrt{B^{2}-C^{2}} ,\\
\tilde{D} &:=& \sqrt{C^{2}-B^{2}}.
\end{eqnarray}
Finally, the domains are $\mathcal{D}_{1} := (-\infty,FL-F\bar{x}_{j}-mc^{2}] \cup [FL-F\bar{x}_{j}+mc^{2},\infty)$ and $\mathcal{D}_{2} := (FL-F\bar{x}_{j}-mc^{2},FL-F\bar{x}_{j}+mc^{2})$. With these results, it is possible to evaluate the wave function numerically in a constant electric field. Note also that this method can be easily adapted to any time-independent potential.

Adapting this numerical method to find $\tilde{\psi}(-L)$ and adding the nuclei, we get that
\begin{eqnarray}
\tilde{\psi}(-L) &=& U(-L,x_{n})U(x_{n},x_{n-1}) \cdots U(x_{j_{N_{p}}+1},x_{j_{N_{p}}})G^{-1}U(x_{j_{N_{p}}},x_{j_{N_{p}}-1}) \nonumber \\
&& \cdots U(x_{j_{1}+1},x_{j_{1}})G^{-1}U(x_{j_{1}},x_{j_{1}-1}) \cdots U(x_{1},L)\tilde{\psi}(L),
\end{eqnarray}
where $\tilde{\psi}(L)$ is given in Eq. \eqref{eq:trans2} and where $x_{j} = L-j \delta x$. This allows us to determine the transmission coefficient by using Eq. \eqref{eq:A_coeff_def}.

\section{Results}
\label{sec:res}

The numerical results obtained from the procedure explained in preceding sections is now presented. In the first part, the case of two nuclei is treated in details, allowing to understand the main features of REPP and ECEPP. Then, the other cases are presented to understand how the pair production rate depends on the field strength and the number of nuclei.

\subsection{Position of resonances and pair production for the two nuclei system}
\label{app:res_pair}

\begin{figure}
\subfloat[]{
\begin{overpic}[scale=.5,unit=1\textwidth]
{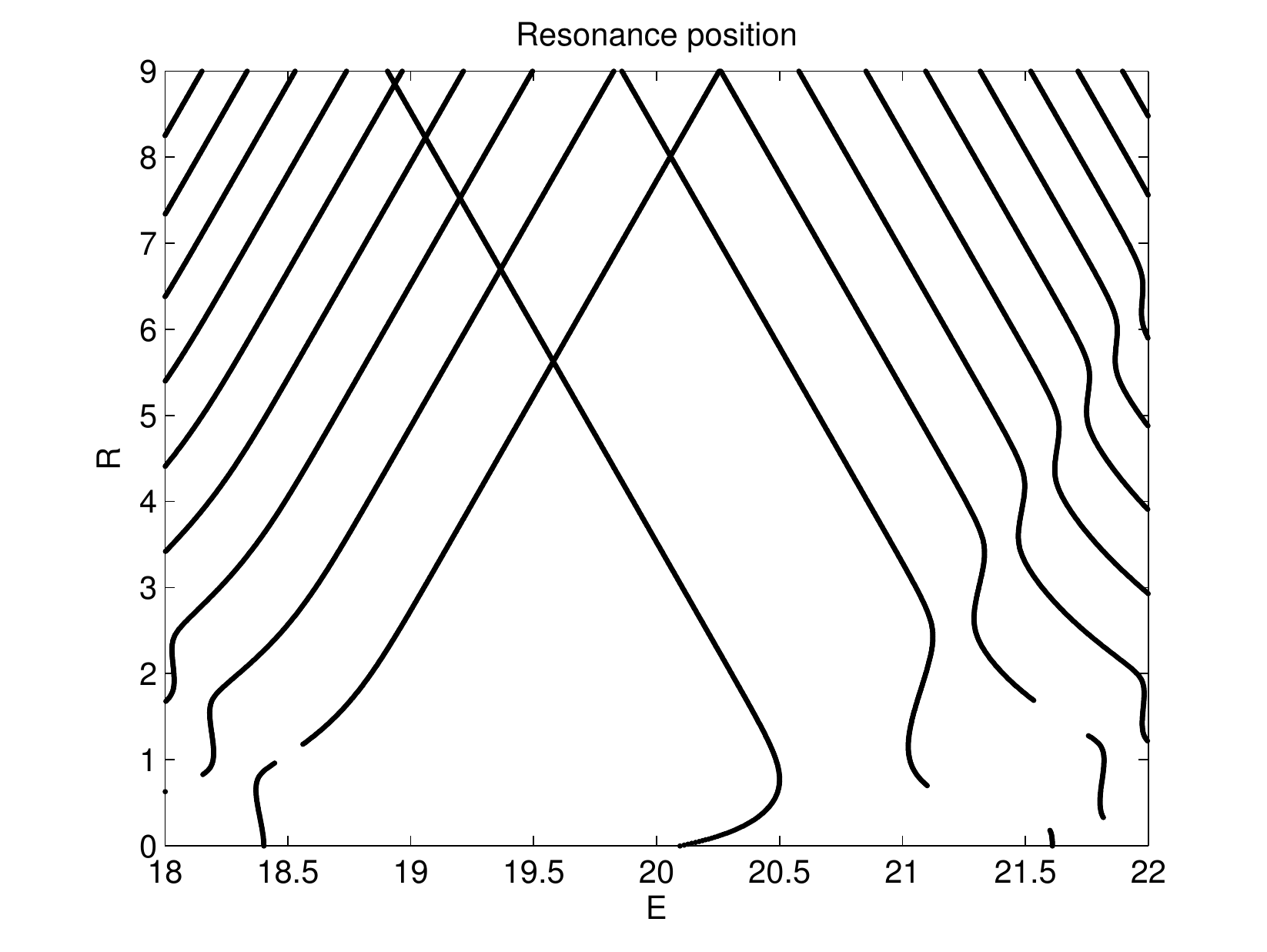}
\put(38,15){\tiny Ground state $\rightarrow$}
\put(42,45){\tikzcircle{0.15cm}}
\put(37.75,52.25){\tikzcircle{0.15cm}}
\put(34.5,58){\tikzcircle{0.15cm}}
\put(31.5,63.5){\tikzcircle{0.15cm}}
\put(51,62){\tikzcirclered{0.15cm}}
\put(61,24){\tiny Excited }
\put(61,22){\tiny state $\rightarrow$}
\put(65,12){
\begin{tikzpicture}
\begin{scope}[color=gray,line width=2pt]
\draw (0,0) -- (2,3.2);
\end{scope}
\end{tikzpicture}
}
\end{overpic}
}
\subfloat[]{
\begin{overpic}[scale=.5,unit=1\textwidth]
{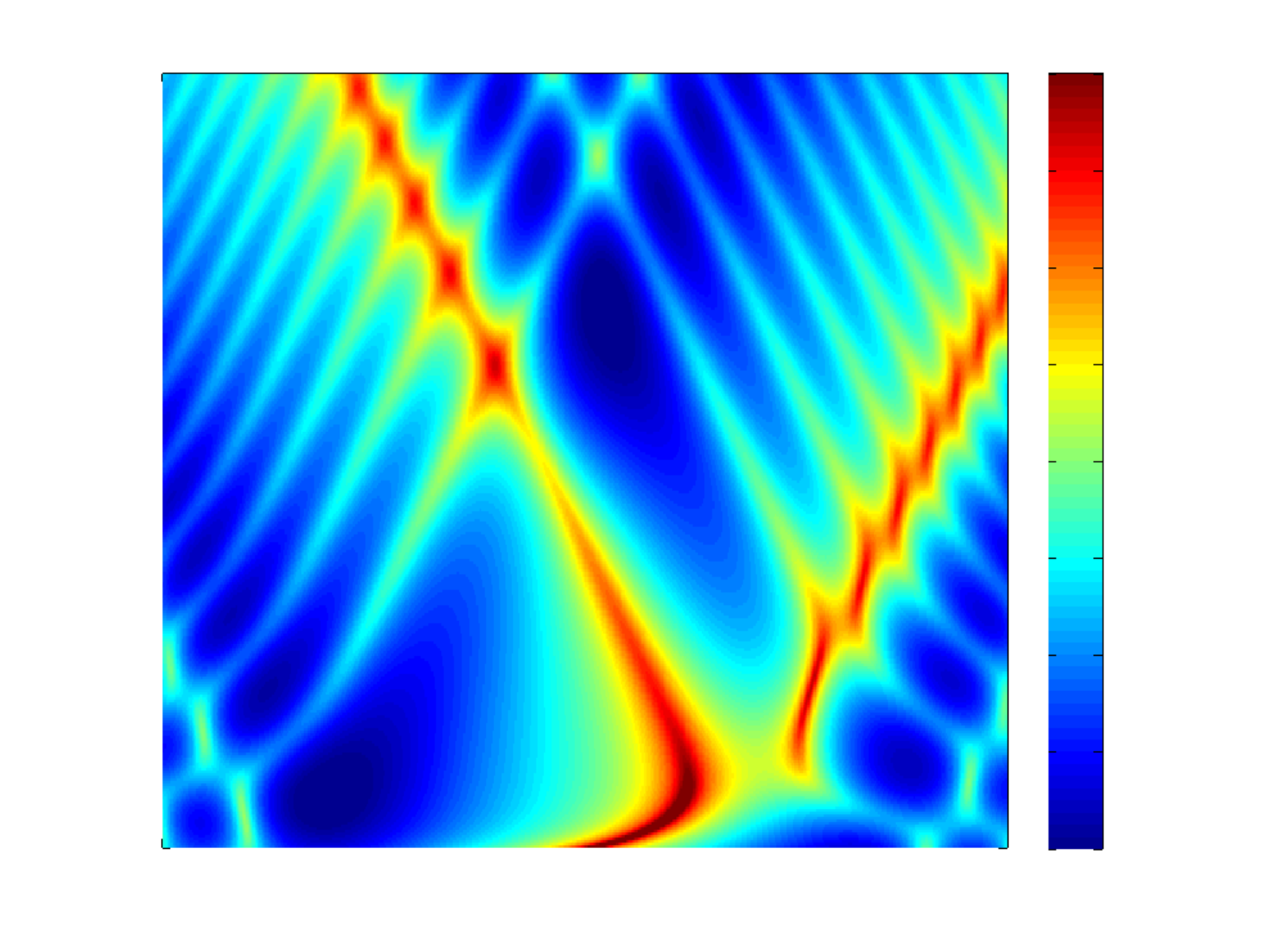}
\put(0,0){\includegraphics[scale=.5]%
{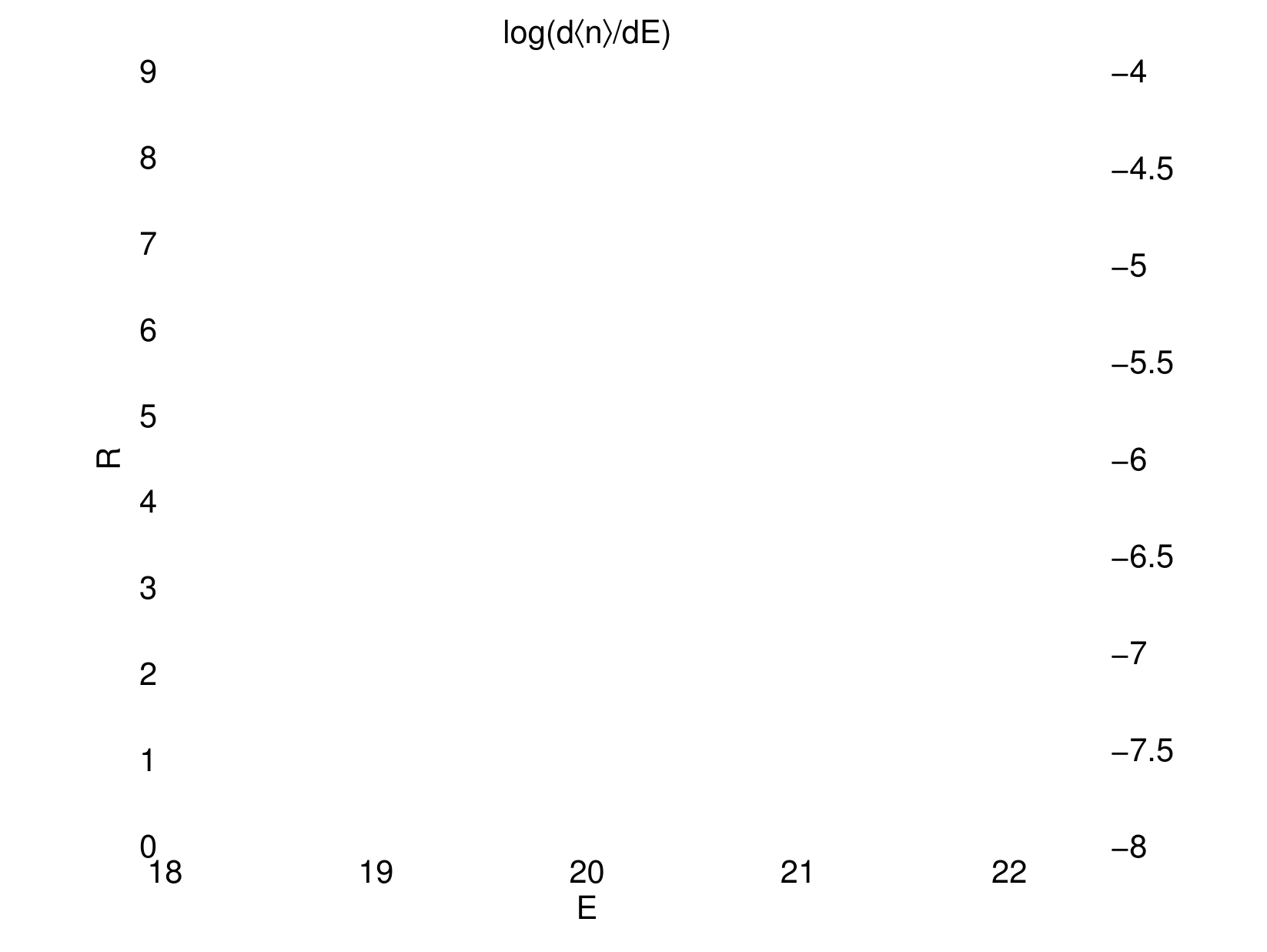}}
\end{overpic}
} \\
\begin{center}
\subfloat[]{\includegraphics[width=0.5\textwidth]{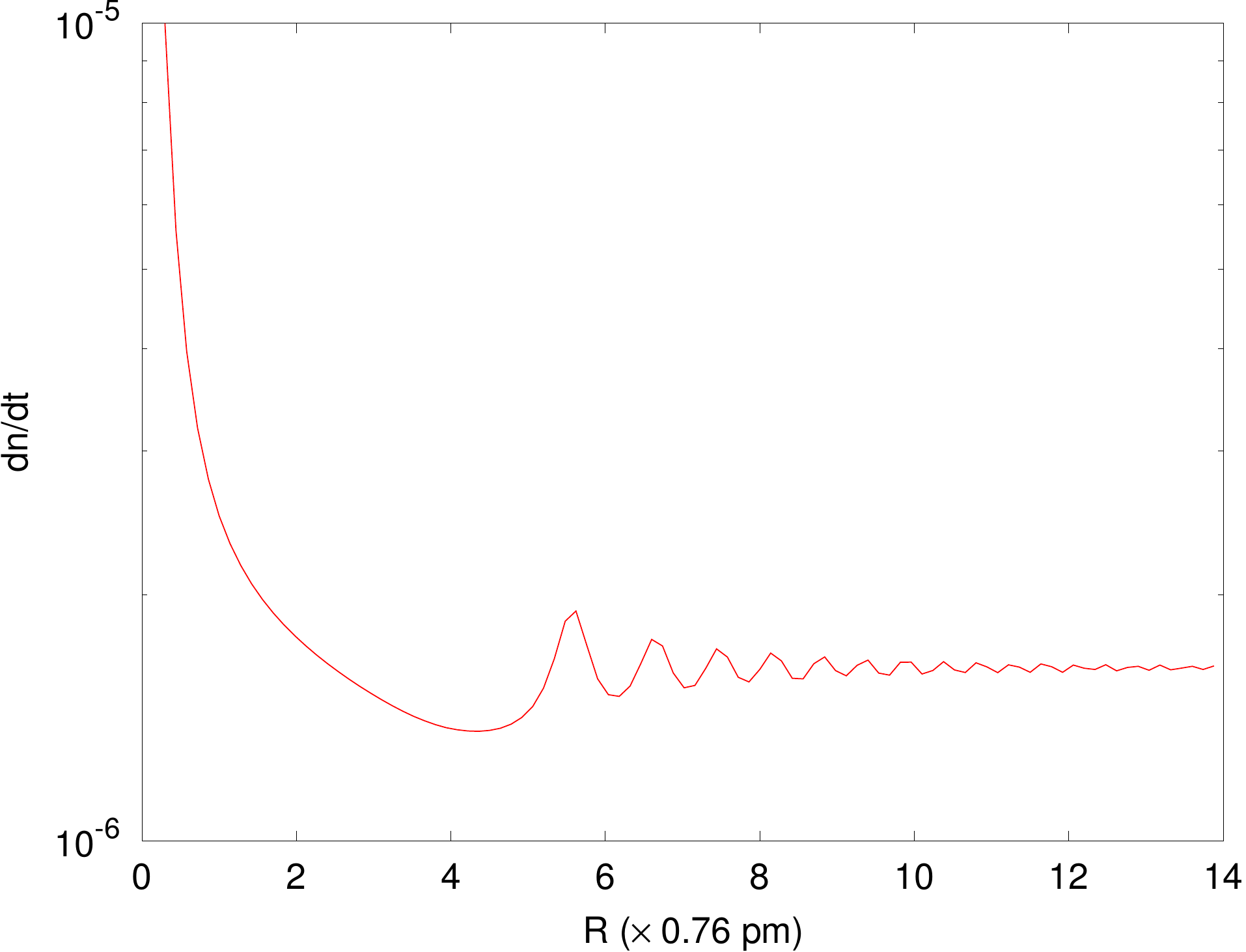}}
\end{center}
\caption{The real part of resonances is shown in (a). The ground state crosses at the blue circles with resonances coming from the negative energy states (channel 1). The excited state follows the grey line and goes through a series of avoided crossings with the resonances coming from the positive energy states (channel 2). The resonances from both continua cross at the red circle (channel 3). The particle spectrum $d\langle n \rangle/dEdt$ as a function of the internuclei distance is shown in (b). There is clearly an enhancement of pair production at the resonance crossings. Finally, in (c), the total rate $d\langle n \rangle/dEdt$ is shown. There are peaks in the pair production rate when the ground state crosses with negative energy continua resonances. The parameters are chosen as $g=0.8$ (corresponding to U$^{91+}$), $F=0.2 \times E_{S}$ and $L=38$ pm. Also, for this figure only, $R$ is the semi-internuclei distance.}
\label{fig:dnde_posi}
\end{figure}
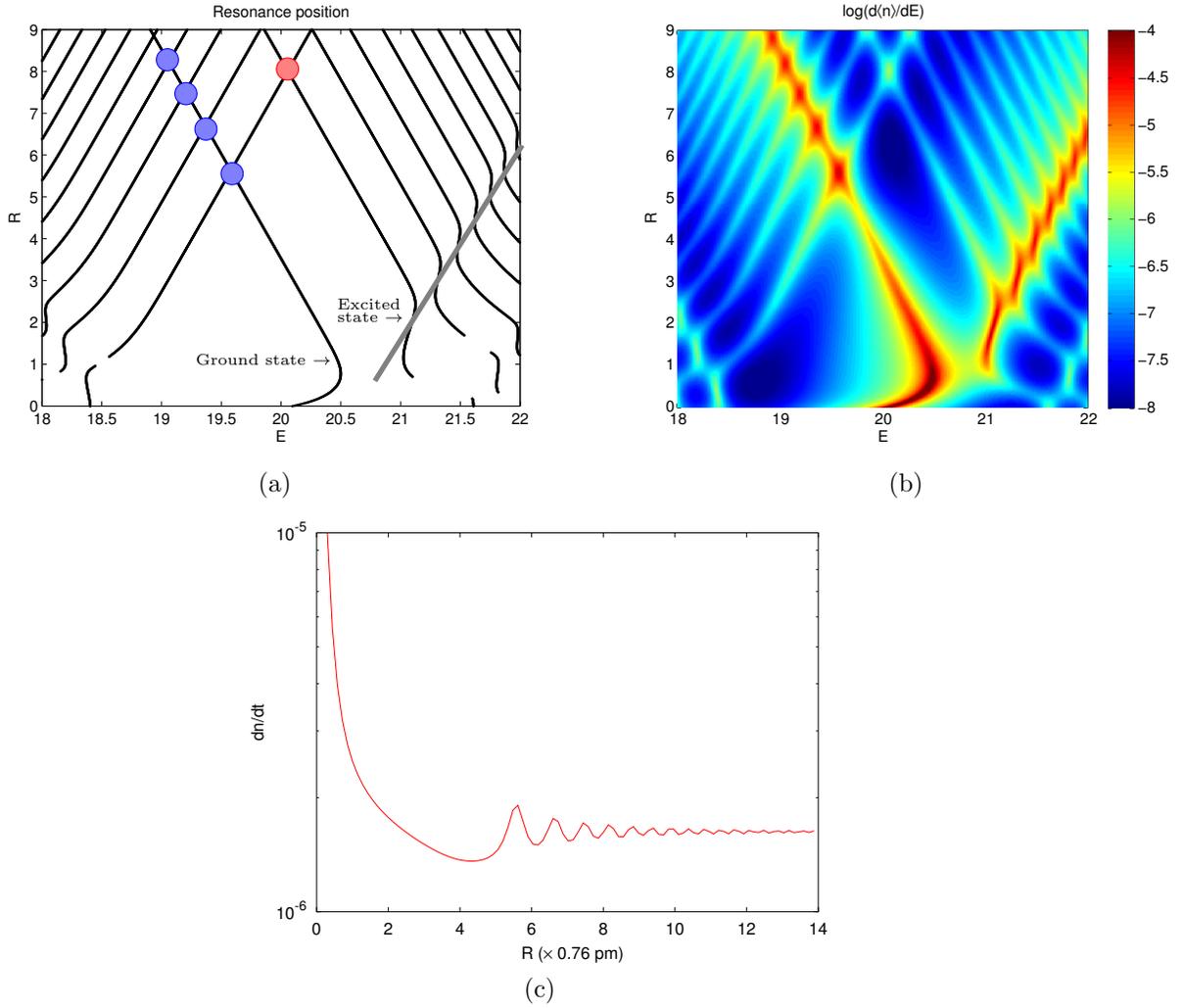

In this section, numerical results are presented for the two nuclei system. The main goal is to show that pair production occurs mainly at the crossing of resonances for larger internuclei distance. The position of resonances can be evaluated by using the Weyl-Titchmarsh-Kodaira (WTK) theory for singular operator on infinite domain \cite{1962eeaw.bookT}. This mathematical method was generalized to the 1-D Dirac equation in \cite{TITCHMARSH01011961,Titchmarsh01011961v}. Within this framework, it is possible to evaluate the spectral density $\rho(E)$ which contain all the information on the spectrum of the operator under study: bound states appear as poles of $\rho$, continua occurs when $\rho$ is analytic and non-zero, and resonances are poles of $\rho$ in the complex energy plane. Using these facts, along with the analytical solution presented in \ref{app:an_sol} and the WTK method, it is possible to obtain an equation giving the position of resonances (for more details, see \cite{1751-8121-45-21-215304}). The latter can be solved numerically and the results are shown in Fig. \ref{fig:dnde_posi}, as a function of the internuclei distance, for the real part of resonance energies. The pair production spectrum, obtained from the transmission-reflection problem, is also shown in this figure. By looking at these two pictures, it is clear that the spectrum $d\langle n \rangle/dtdE$ is enhanced when the resonances are crossing (when two lines are crossing in (a)). For instance, there is a peak in the spectrum at $E\approx 19.5 \; (\times mc^{2})$ and $R\approx 5.5 \; (\times 0.76 \; \mbox{pm})$, where the ground state resonance crosses with a resonance coming from the negative energy states. From these figures, it is possible to conclude that pair production proceeds via three channels \cite{PhysRevLett.110.013002}:
\begin{itemize}
\item Channel 1: the ground state crosses with resonances coming from the negative energy states (blue circles in the figure).
\item Channel 2: the excited state goes through avoided crossings with resonances coming from the positive energy continuum.
\item Channel 3: resonances from the negative energy continuum crosses with resonances from the positive continuum (red circle in the figure).
\end{itemize}
The physical interpretation of each channel is the following. For channel 1, it is possible for a negative energy state to tunnel to the ground state at the crossing and this enhances pair production. As shown in Fig. \ref{fig:dnde_posi}(c) for the total rate, it is clearly the dominant process: at the crossing, a peak in the rate appears. For channel 2, the enhancement is due to the ionization of the excited state: when it crosses with resonances, the ionization is enhanced and this reduces Pauli blocking. Thus, a transition from the negative energy states to the excited state is possible. Finally, for channel 3, there is a direct transition between negative and positive energy states. Clearly, channel 2 and 3 are not as important as channel 1 as they are not leading to any significant effects in the total rate (there are no visible peaks due to these crossings).

\subsection{Total rate for many-center systems}

\begin{figure}
\subfloat[]{\includegraphics[width=0.5\textwidth]{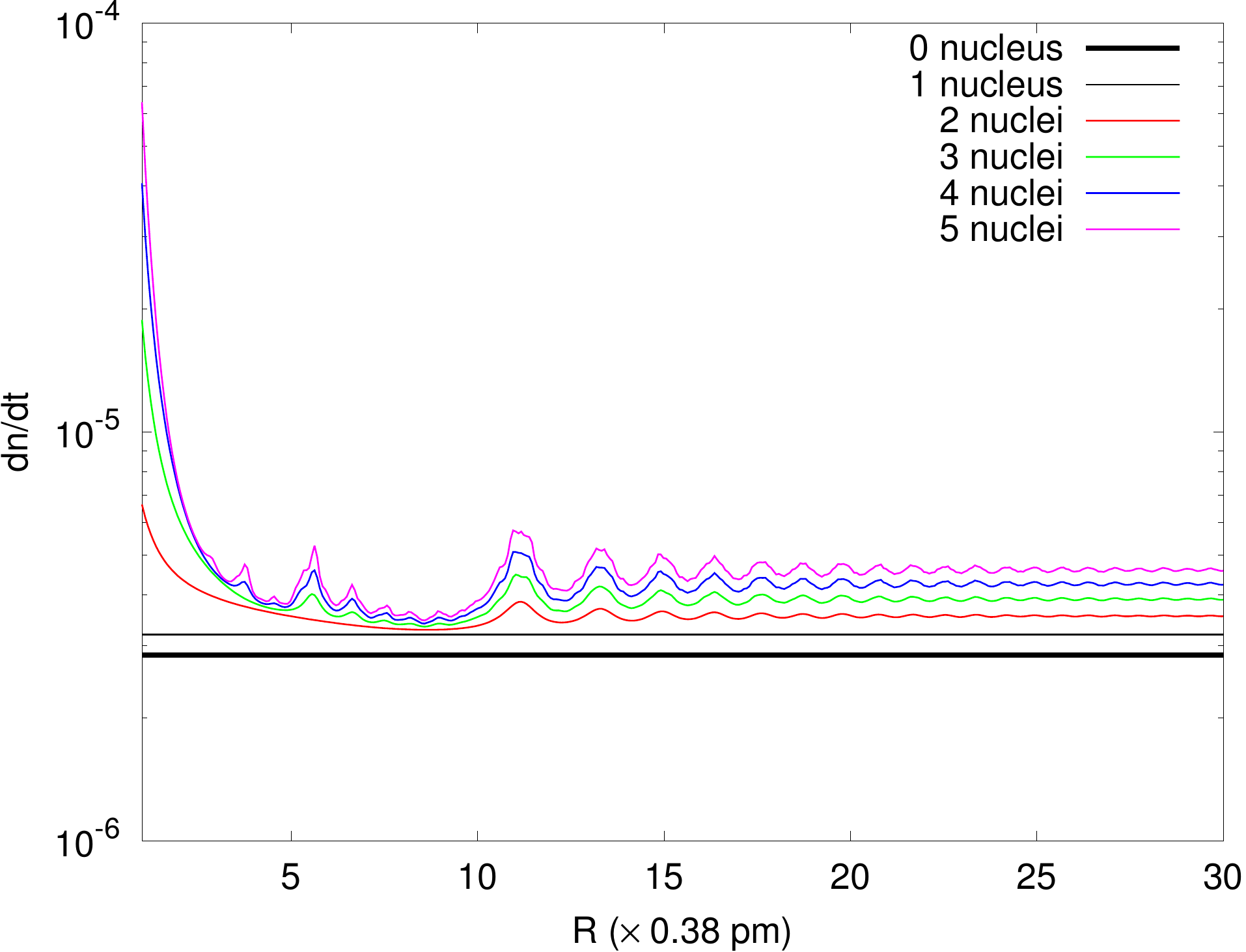}}
\subfloat[]{\includegraphics[width=0.5\textwidth]{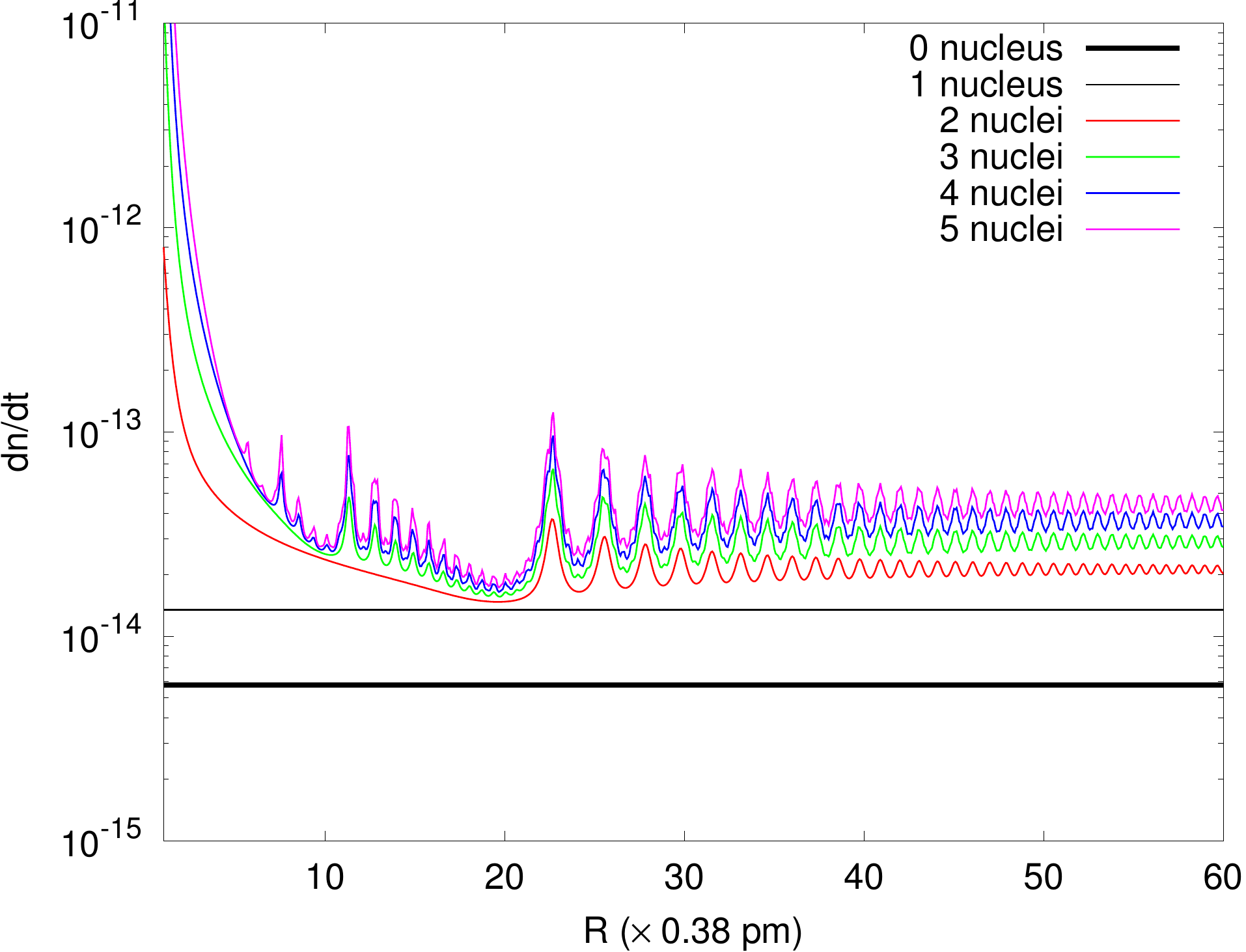}} \\
\begin{center}
\subfloat[]{\includegraphics[width=0.5\textwidth]{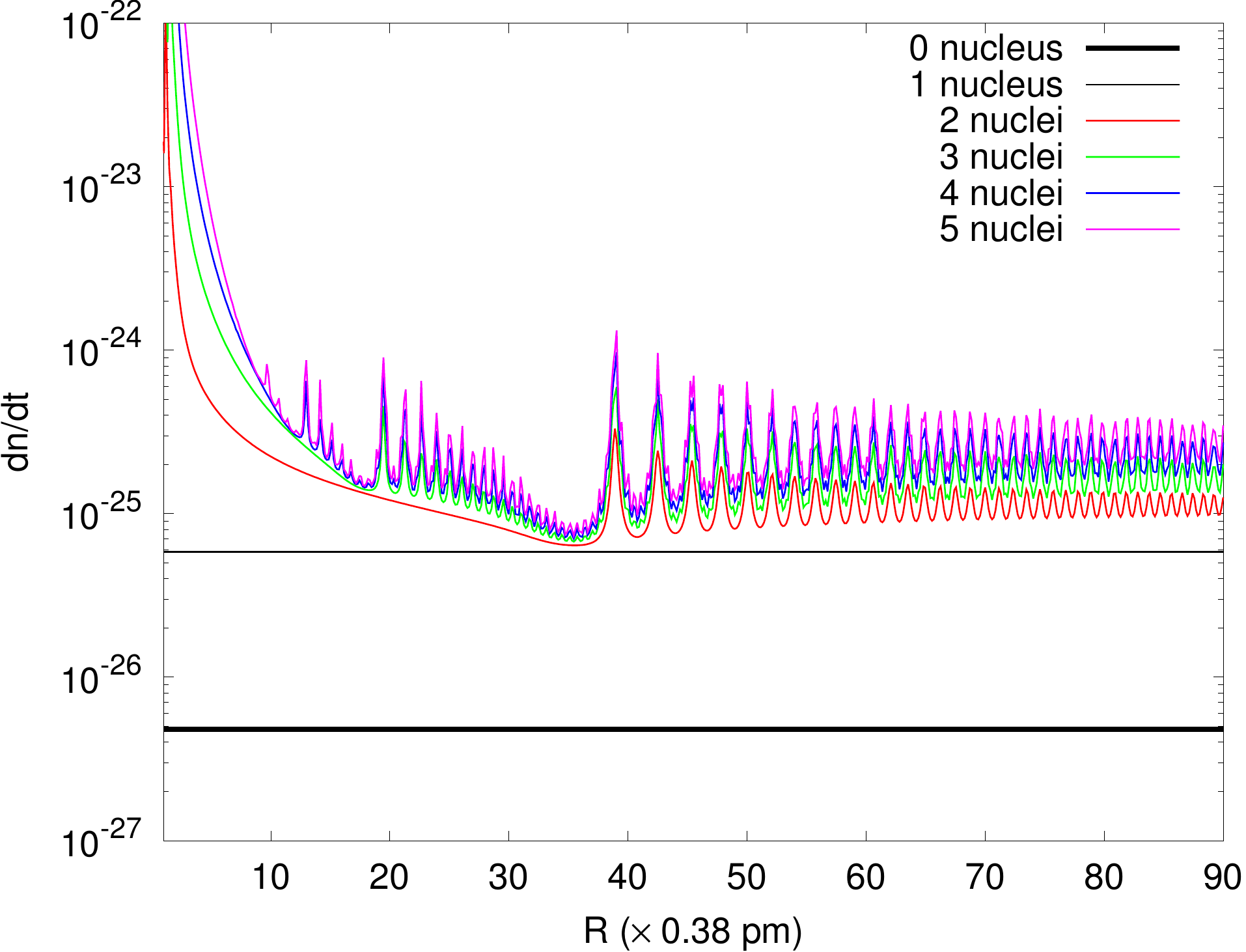}}
\end{center}
\caption{Total rate for 0, 1, 2, 3, 4 and 5 nuclei, as a function of internuclei distance $R$, for (a) $F = 0.2 \times E_{S}$, (b) $F = 0.09 \times E_{S}$ and (c) $F = 0.05 \times E_{S}$. The strength of the potential well is set to $g = 0.8$ (corresponding to Uranium nuclei).}
\label{fig:dndt_many}
\end{figure}

\begin{figure}
\begin{center}
\begin{overpic}[scale=.8,unit=1\textwidth]
{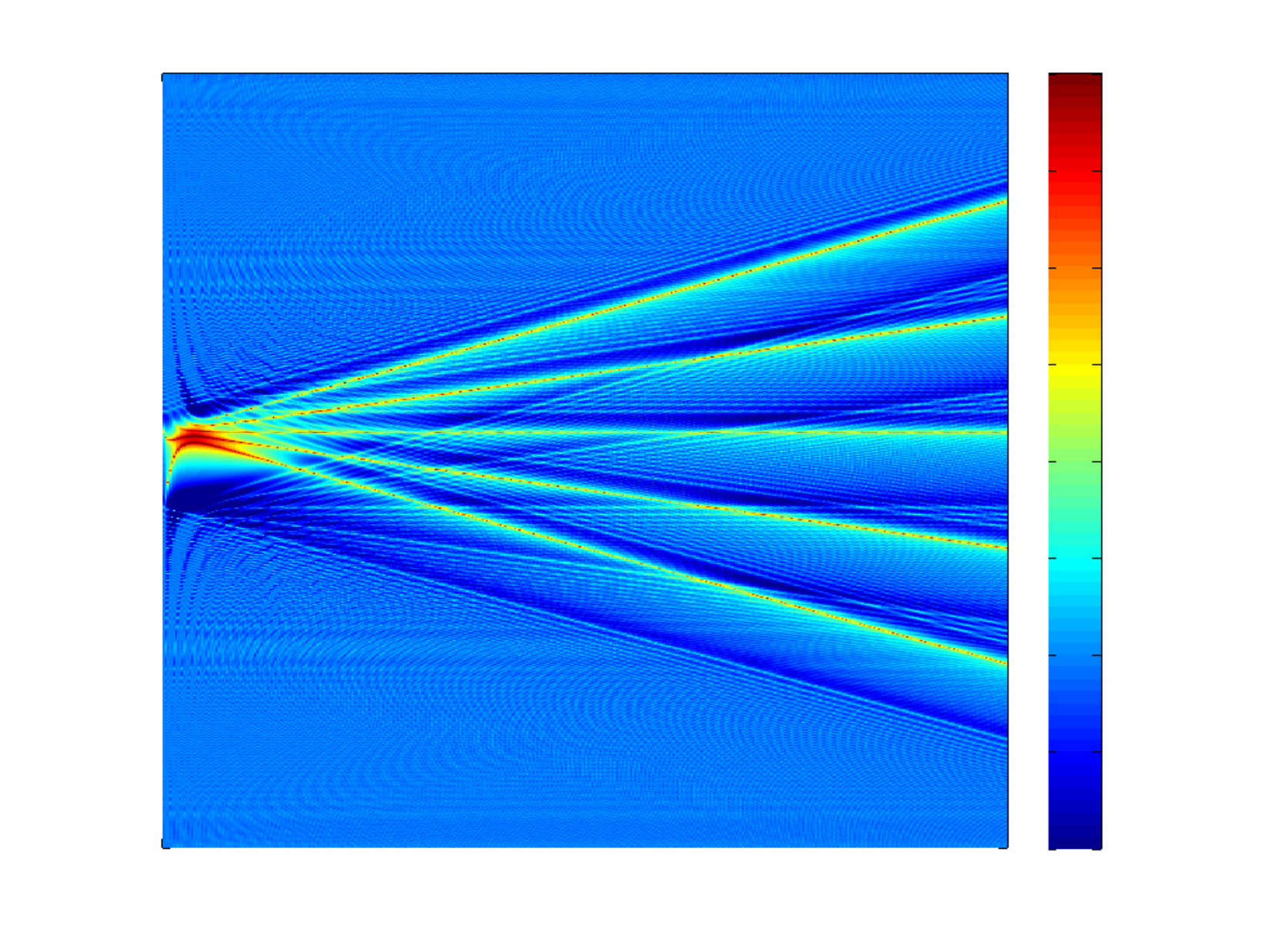}
\put(0,0){\includegraphics[scale=.8]%
{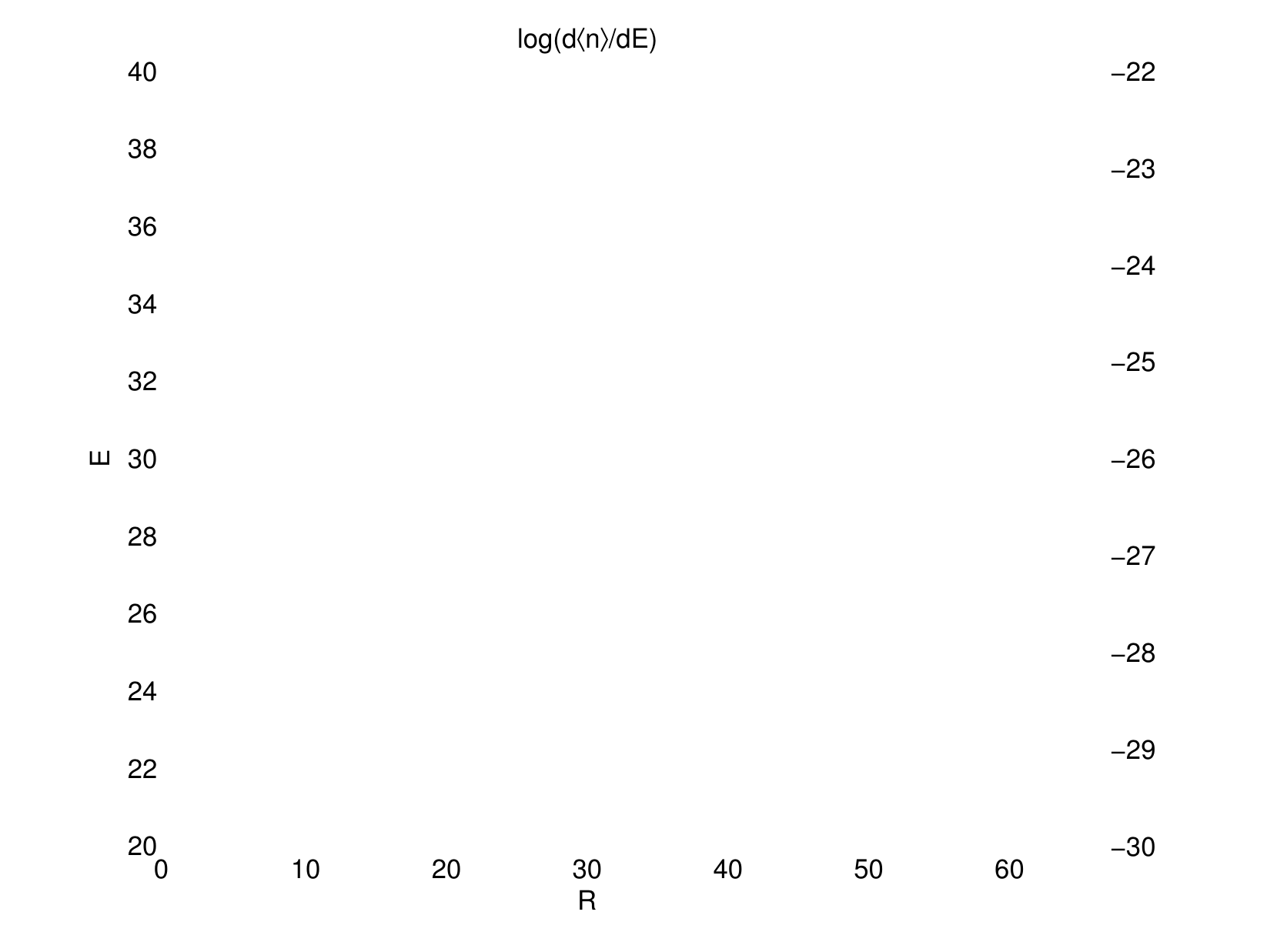}}
\end{overpic}
\end{center}
\caption{Pair production spectrum for 5 nuclei. The Stark shift of resonances can be easily seen. Many crossings occurs in this case as the resonance structures is much more intricate than in the two-center case. }
\label{fig:dnde_5nuc}
\end{figure}

The total rates obtained numerically are depicted in Fig. \ref{fig:dndt_many} for 0 to 5 nuclei, as a function of the internuclei distance. In all cases, the largest enhancement occurs when $R$ is small, which corresponds to the ECEPP mechanism. At larger $R$, there is a peak structure that emerges which comes from the crossing of resonances (REPP mechanism). This structure depends on the number of nuclei: for instance, for a larger number of nuclei, there are new structures appearing at smaller internuclei distance (this can be seen more clearly in Fig. \ref{fig:dndt_many} (c), for $F=0.05 \times E_{S}$). This is due to the fact the number of quasi-degenerate ground states equals the number of nuclei. In the electric field, each of these state is Stark shifted. Then, the number of crossings is increased as many states can cross with negative energy resonances, producing the peak structure in the rate. This is shown more explicitly in the pair production spectrum for the 5 nuclei case, in Fig \ref{fig:dnde_5nuc}.

There is also a few important remarks concerning the rates of many-center systems:
\begin{enumerate}
\item The relative enhancement (with respect to the 0 and 1 nucleus cases) increases as the electric field strength is reduced. This is related to the stability of resonances: when the electric field is smaller, these states are more stable and their decay rate (related to the inverse of the imaginary part of the eigenenergy) increases. Thus, the transition between these states is enhanced when they are crossing.   
\item The largest peak occurs at a larger internuclei distance when the electric field is reduced. This is related to the strength of the Stark shift which scales linearly with the electric field strength.  
\end{enumerate}

\section{Conclusion}
\label{sec:conclu}

In this work, the electron-positron production rate was evaluated for many-center systems. It was shown that two mechanisms can enhance the rate: the ECEPP and the REPP. On the one hand, at small internuclei distance, the enhancement is due to the ECEPP. In this case, the Coulomb force becomes more important than the constant electric field and the bound states resonances have low energy, close to the negative energy states. Thus, it is easier for an electron in the negative energy sea to tunnel into these resonances. On the other hand, at larger internuclei distance, the enhancement is due to the REPP. In this regime, the constant electric field is more important than the Coulomb forces of the nuclei and the Stark shift of resonances becomes the dominant effect: the negative energy resonances are shifted up in the spectrum while the ground states is shifted down. When the resonances from the negative energy states cross with the resonances of the positive energy states, there is an enhancement of the pair production rate. 

The two mechanisms were observed for every number of nuclei considered (except zero and one nucleus case, of course). It was demonstrated that as the number of nuclei is increased, the peak structure becomes more intricate. This is related to the fact that the number of quasi-degenerate ground states is given by  the number of nuclei. As these are Stark shifted, there are new crossings of these quasi-degenerate ground state with negative energy states and this induces new peak structures in the production rate. The height of peaks existing at a smaller number of nuclei is also enhanced and can reach a few orders of magnitude above Schwinger's mechanism. This effect (REPP) may be useful in the study of the Dirac vacuum in laser-matter interaction experiments as it does not require small internuclei distance (as in ECEPP), which are only achievable in relativistic heavy ion collisions. 

In the more realistic case of 3D nuclei modelled by Coulomb potentials, similar effects should be observed but the spectrum of the Coulomb potential contains an infinite number of states, multiplying the possible crossings of resonances. Nevertheless, REPP will proceed in a similar way as in the simple model presented in this article because the behavior of the ground state resonance is qualitatively the same in both approaches. Therefore, it is expected that the peak structures will be only slightly modified. However, in 3D, the negative energy state resonances may be less stable than in 1D and this may have an effect on the width and height of the peaks in the total rate. This is presently under investigation.

\ack
The authors would like to thank S. Chelkowski for numerous discussions on REPP and other topics.

\appendix

\section{Analytical solution}
\label{app:an_sol}

The Dirac equation with the constant electric field of Eq. \eqref{eq:elec_pot} can be solved analytically by decoupling the two spinor components and by letting $ y(x) = e^{-i\frac{\pi}{4}}\sqrt{\frac{2c}{F}} \left( \frac{E-F(x-L)}{c} \right)$. 
%
%
Then, the Dirac equation becomes a system of equations with well-known solutions in terms of parabolic cylinder functions $U(\gamma,z)$ \cite{DLMF}:
\begin{eqnarray}
 \psi(x) &=& c_{1} U_{a}(x) + c_{2} U_{b}(x),
\end{eqnarray}
%
where $c_{1,2}$ are integration constants and where we defined
\begin{eqnarray}
U_{a,1}(x)&:= & U(\gamma,y(x)) ,\\ 
U_{b,1}(x) &:= &  U(-\gamma,-iy(x)), \\
U_{a,2}(x)&:= &mc \sqrt{\frac{c}{2F}} e^{i\frac{3\pi}{4}} U(\gamma+1,y(x)) ,\\
U_{b,2}(x) &:= & \frac{1}{mc} \sqrt{\frac{2F}{c}} e^{-i\frac{\pi}{4}} U(-\gamma-1,y(x)) .
\end{eqnarray}
Here, we have $\gamma = i\frac{m^2 c^3}{2F} - \frac{1}{2}$.

\section*{References}
\bibliographystyle{iopart-num}
\bibliography{bibliography}


\end{document}